\documentclass[aps,prx,reprint,twocolumn,superscriptaddress,showpacs,amssymb]{revtex4-1}
\usepackage[english]{babel}
\usepackage{graphics}
\usepackage{graphicx}
\usepackage{epsfig}
\usepackage{amssymb}
\usepackage{dcolumn}
\usepackage{bm}
\usepackage{color}
\usepackage{natbib}
\usepackage{lineno}
\usepackage{sidecap}
\usepackage{verbatim}  
\usepackage{vector}  
\usepackage{wrapfig}
\usepackage{enumerate}
\usepackage{sidecap}
\usepackage{amsmath}
\usepackage{hyperref}

\begin{document}


\title{Metal-chalcogen bond-length induced electronic phase transition from semiconductor to topological semimetal in ZrX$_2$ (X = Se and Te)}

\author{I.\ Kar}
\affiliation{Condensed Matter Physics and Material Science Department, S N Bose National Centre for Basic Sciences, 700106, India}
\author{Joydeep Chatterjee}
\affiliation{Condensed Matter Physics and Material Science Department, S N Bose National Centre for Basic Sciences, 700106, India}
\author{ Luminita Harnagea}
\affiliation{Indian Institute of Science Education and Research, Dr. Homi Bhabha Road, Pune, Maharashtra-411008, India}
\author{Y. Kushnirenko}
\affiliation{Leibniz Institute for Solid State Research, IFW Dresden, D-01171 Dresden, Germany}
\author{ A. V. Fedorov}
\affiliation{Leibniz Institute for Solid State Research, IFW Dresden, D-01171 Dresden, Germany}
\author{Deepika Shrivastava}
\affiliation{Condensed Matter Physics and Material Science Department, S N Bose National Centre for Basic Sciences, 700106, India}
\author{B. B\"uchner}
\affiliation{Leibniz Institute for Solid State Research, IFW Dresden, D-01171 Dresden, Germany}
\author{P. Mahadevan}
\affiliation{Condensed Matter Physics and Material Science Department, S N Bose National Centre for Basic Sciences, 700106, India}
\author{S.\ Thirupathaiah}
\email{setti@bose.res.in}
\affiliation{Condensed Matter Physics and Material Science Department, S N Bose National Centre for Basic Sciences, 700106, India}
\affiliation{Leibniz Institute for Solid State Research, IFW Dresden, D-01171 Dresden, Germany}
\date{\today}

\begin{abstract}
Using angle resolved photoemission spectroscopy (ARPES) and density functional theory (DFT)  calculations  we studied the low-energy electronic structure of bulk ZrTe$_2$. ARPES studies on ZrTe$_2$ demonstrate free charge carriers at the Fermi level, which is further confirmed by the DFT calculations. An equal number of hole and electron carrier density estimated from the ARPES data, points ZrTe$_2$ to a semimetal. The DFT calculations further suggest a band inversion between Te $p$ and Zr $d$ states at the $\Gamma$ point, hinting at the non-trivial band topology in ZrTe$_2$. Thus, our studies for the first time unambiguously demonstrate that ZrTe$_2$ is a topological semimetal.  Also, a comparative band structure study is done on ZrSe$_2$  which shows a semiconducting nature of the electronic structure with an indirect band gap of 0.9 eV between $\Gamma (A) $ and $M (L)$ high symmetry points.  In the below we show that the metal-chalcogen bond-length plays a critical role in the electronic phase transition from semiconductor to a topological semimetal ingoing from ZrSe$_2$ to ZrTe$_2$.
\end{abstract}
\maketitle

\section{Introduction}
Transition metal dichalcogenides (TMDCs)~\cite{Dickinson1923, Wilson1969, Mattheiss1973} are gaining a great deal of research interests, especially,  from the past few decades due to their potential applications in the spintronics~\cite{Spaeh1983, Jimenez2012, Wang2012d, Zeng2012,Cheng2014, Cheng2014} and as well in the optoelectronics~\cite{Li1996, Fuhrer2013, Wu2014, Gong2017, Khan2018}  because of their wide range of electronic properties  starting from the metallic~\cite{Neal2014, Zhang2017, Zhang2017a}, to the semimetallic~\cite{Ali2014, Jiang2017, Thirupathaiah2017, Ma2018},  to the semiconducting~\cite{Zheng1989, Lin2014, Hill2016, Salavati2018}, and to the Mott-insulators~\cite{Pillo2000,  Perfetti2006, Zhang2016a},  obtained mainly by the band engineering ~\cite{Mak2010,Kumar2012a, Choi2017}. In addition, the diversity of electronic properties of TMDCs includes the charge density wave (CDW)~\cite{Porer2014, Li2016a}, the magnetism~\cite{Ma2012, Zhu2016, Xiang2016}, and the superconductivity~\cite{Joe2014, Liu2017b, Barrera2018}.

Tunable band gaps in layered TMDCs is one of the major research topics in recent days from both the theory and experiment~\cite{Ramasubramaniam2011, Das2014, Movva2018, Aghajanian2018}. It is known that the band gaps in IVB TMDCs (TiX$_2$, ZrX$_2$ and HfX$_2$, where X = S, Se, and Te) increase with the strain~\cite{Guo2014a}. However, among the IVB TMDCs,  ZrX$_2$ compounds in their monolayer thickness show peculiarity beyond a critical pressure. That means,  instead of an increase in the band gap with the pressure, it starts decreasing with increasing the pressure. On the other hand, the monolayer thickness ZrTe$_2$ transforms from metallic to a semiconductor beyond a critical pressure.  Hence, understanding the pressure effects on electronic structure of TMDCs has a significance in their band engineering~\cite{Feng2014a}.  Specifically, performing ARPES studies under external pressure is a non-trivial method~\cite{Yi2011a, Zhang2012a}. However, studying the electronic structure using ARPES under the application of chemical pressure induced by the substitution of an isovalent atom is highly viable~\cite{Thirupathaiah2011}. So far there exists very few ARPES reports in this direction~\cite{Brauer1995, Moustafa2009, Moustafa2013, Mleczko2017, Ghafari2018}. Moreover, these studies report only the electronic structure of Zr(Se$_{1-x}$S$_x$)$_2$ which are semiconductors. Although several theoretical reports on ZrTe$_2$ predicted it to be a metal~\cite{Reshak2004, Kumar2015, Machado2017}, so far no ARPES study exists on this compound in bulk phase confirming the same,  except that a recent ARPES study on monolayer ZrTe$_2$ deposited on InAs (111) substrate showing linear Dirac states near the Fermi level at the $\Gamma$-point~\cite{Tsipas2018}. Thus, a thorough understanding of the electronic structure of ZrTe$_2$ has other vested interests as well,  whether ZrTe$_2$ is another topological system like ZrTe$_5$~\cite{Whangbo1982,Manzoni2016,Zhang2017b, Shen2017, Xiong2017}. Therefore, ARPES studies on the bulk ZrTe$_2$ are essential to confirm whether ZrTe$_2$ is a simple metal or a topological semimetal as suggested by a recent transport study~\cite{Machado2017}.

\begin{figure*}[htbp]
  \centering
     \includegraphics[width=0.98\textwidth]{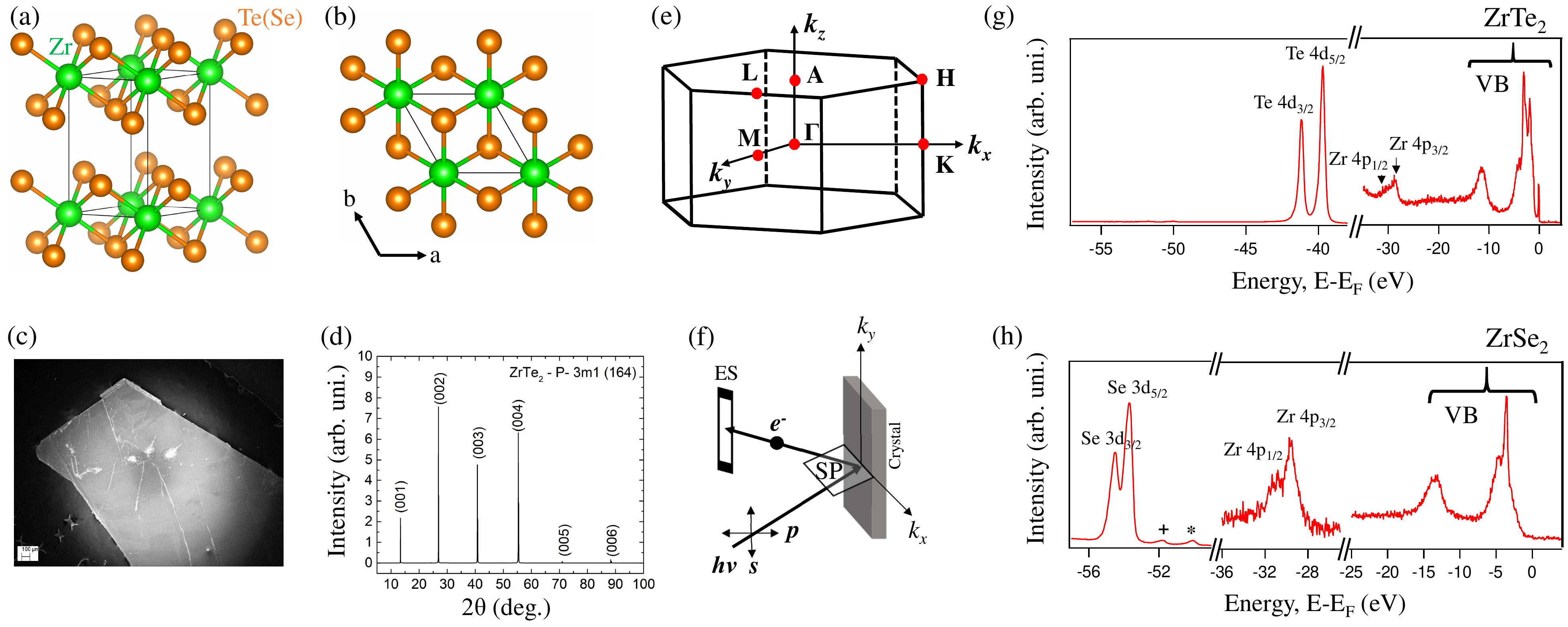}
        \caption{(a) Trigonal crystal structure of ZrTe$_2$ and ZrSe$_2$. (b) ZrX$_2$ layer projected on the $ab$-plane, showing the metal-chalcogen hexagonal pattern. (c) Scanning electron microscope image of the ZrTe$_2$ single crystal with a spacial resolution of 100 $\mu m$. (d) X-ray diffractogram of ZrTe$_2$ single crystal.  (e) 3D view of the hexagonal Brillouin zone with high symmetry points.  (f) Measuring geometry of photoemission spectroscopy, defining the $s$ and $p$-polarized photons with respect to the scattering plane (SP) and analyzer entrance slit (ES). Core level photoemission spectra of (g) ZrTe$_2$ and (h) ZrSe$_2$. In (h), the symbols + and $\star$  represent the Iodine impurity peaks of 4p$_{3/2}$ and 4p$_{5/2}$, respectively.}
        \label{1}
\end{figure*}

In this paper, we report on the low-energy electronic structure of bulk ZrTe$_2$ and ZrSe$_2$ using high resolution angle resolved photoemission spectroscopy and density functional theory. On the Fermi surface map of ZrTe$_2$, we observe several well disconnected hole and electron pockets at $\Gamma (A)$ and $M (L)$ points, respectively.  Further, in ZrTe$_2$, we realize three holelike non-degenerate band dispersions near the $\Gamma (A)$ point and an electronlike band dispersion at the $M (L)$. Our DFT calculations predict all hole pockets are composed by the Te $p$ orbital characters and the electron pocket is composed by the Zr $d_{z^2}$ orbital character. An equal number of hole and electron carrier density estimated from our ARPES data using Luttinger's theorem ~\cite{Luttinger1960, Oshikawa2000} suggest ZrTe$_2$ to a semimetal. In addition, our DFT calculations on ZrTe$_2$ in presence of spin-orbit coupling (SOC) predict band inversion between Te $p$ and Zr $d$ characters near $\Gamma$ point, pointing ZrTe$_2$ to a topological semimetal.  On the other hand, in contrast to ZrTe$_2$,  from the Fermi surface topology of ZrSe$_2$ we observe only electron pockets located at the $M (L)$ point, while the hole pockets are noticed well below the Fermi level at a binding energy of $\approx$ 1 eV. Nevertheless, our studies on ZrSe$_2$ are inline with the existing reports that it is a semiconductor with an indirect band gap of 0.9 eV between $\Gamma (A)$ and $M (L)$ high symmetry points~\cite{ Mleczko2017, Ghafari2018, Nikonov2018}. We further discuss the effect of metal-chalcogen bond length on the electronic structure of ZrX$_2$ ( X = Se and Te) and show that the metal-chalcogen bond length is the source of electronic phase transition of these systems from semiconducting to semimetallic.

\section{EXPERIMENTAL DETAILS}

High quality single crystals of ZrSe$_2$ and ZrTe$_2$ were grown by the chemical vapour transport (CVT) technique using iodine as transporting agent~\cite{Schaefer1964}.  In the first step, stoichiometric quantities of elements, Zr (sponge, ≥ 99.9\%, metals basis, Sigma Aldrich) with Se (shot, Alfa Aesar, 99.999\%, metals basis) or Te (ingot, 99.99\%, metals basis, Alfa Aesar), were loaded into  an alumina crucible and subsequently sealed in a quartz ampoule under vacuum. The ampoule was slowly heated to 500 $^{\circ}$C, with a rate of 50 $^{\circ}$C/h, kept there for 5 h and then heated further to 900 $^{\circ}$C. This temperature has been maintained for a period of 2 days, in order to complete the reaction between Zr and the chalcogen (Se/Te). The ZrSe$_2$ and ZrTe$_2$ polycrystalline samples obtained were further used for the crystal growth. The powdered samples were placed and sealed under vacuum in quartz ampoules together with pieces of crystallized iodine (5 mg/cm$^3$). The ampoules had subsequently been loaded in a three-zone furnace, where a gradient of temperature of 100 $^{\circ}$C was maintained between the source (880 $^{\circ}$C) and the sink (780 $^{\circ}$C) zone. After 10 days of transport, many single crystals with dimensions as large as 5mm $\times$ 5mm $\times$ 0.1mm  were obtained at the cold part of the ampoule.  Crystal structure, morphology and chemical composition of the single crystals were determined using X-ray diffraction  and scanning electron microscope equipped with an energy dispersive X-ray spectroscopy probe (EDX) (see Figure~\ref{1}). Both the compounds crystallized in the space group of P-3m1 (164), with lattice parameters at room temperature as $a$ = $ b$ = 3.945 $\AA$ and $c$ = 6.624 $\AA$ for ZrTe$_2$ and $a$ = $b$ = 3.766 $\AA$ and $c$ = 6.150 $\AA$ for ZrSe$_2$.

Angle resolved photoemission spectroscopy measurements were carried out at the UE-112 beam-line equipped with 1$^3$-ARPES end station located in BESSY II (Helmholtz zentrum Berlin) synchrotron radiation center~\cite{Borisenko2012a, Borisenko2012b}. Photon energies for the measurements were varied between 30 eV to 110 eV. The energy resolution was set between 10 and 15 meV depending on the excitation energy. Data were recorded at a chamber vacuum of the order of 1 $\times$ 10$^{-10}$ mbar and the sample  temperature was kept at 1 K during the measurements.  We employed various photon polarizations in order to extract the electronic structure comprehensively.

\section{BAND STRUCTURE CALCULATIONS}

The electronic structure of ZrSe$_2$ and ZrTe$_2$ have been calculated within a projected augmented plane wave (PAW) method of density functional theory as implemented in the Vienna Ab-initio Simulation Package (VASP)~\cite{Kresse1996, Kresse1999}. A k-mesh of 16 $\times$ 16 $\times$ 16 was considered for the k-point integration while plane-wave cutoff 700 eV was used for the basis sets. The GGA-PBE approximation to the exchange correlation functional was used~\cite{Perdew1996}. The experimental crystal structures were taken for the starting structural information. A full optimization of the lattice parameters as well as the internal positions was carried out. The optimized in-plane lattice parameter for ZrSe$_2$ was found to be $a$ = $b$ = 3.735 {\AA} and that along the stacking direction $c$ = 6.206 {\AA}. Similarly for ZrTe$_2$, the optimized lattice parameters along the in-plane direction $a$ = $b$ =  3.909 {\AA} and that along the out-of-plane direction $c$ = 6.749 {\AA} were obtained. They represent about 1-2 {\%} deviations from the  experimental values. In order to quantify the electronic structure changes between ZrTe$_2$ and ZrSe$_2$, we carried out a DFT band mapping  onto a tight-binding model using the VASP to Wannier90 interface ~\cite{Mostofi2008, Franchini2012} which had maximally localized Wannier function for the radial parts of the wave function. Zr $d$ and Se/Te $s$ and $p$ states were included for the band mapping.

\section{RESULTS}

\begin{figure}
  \centering
  \includegraphics[width=0.49\textwidth]{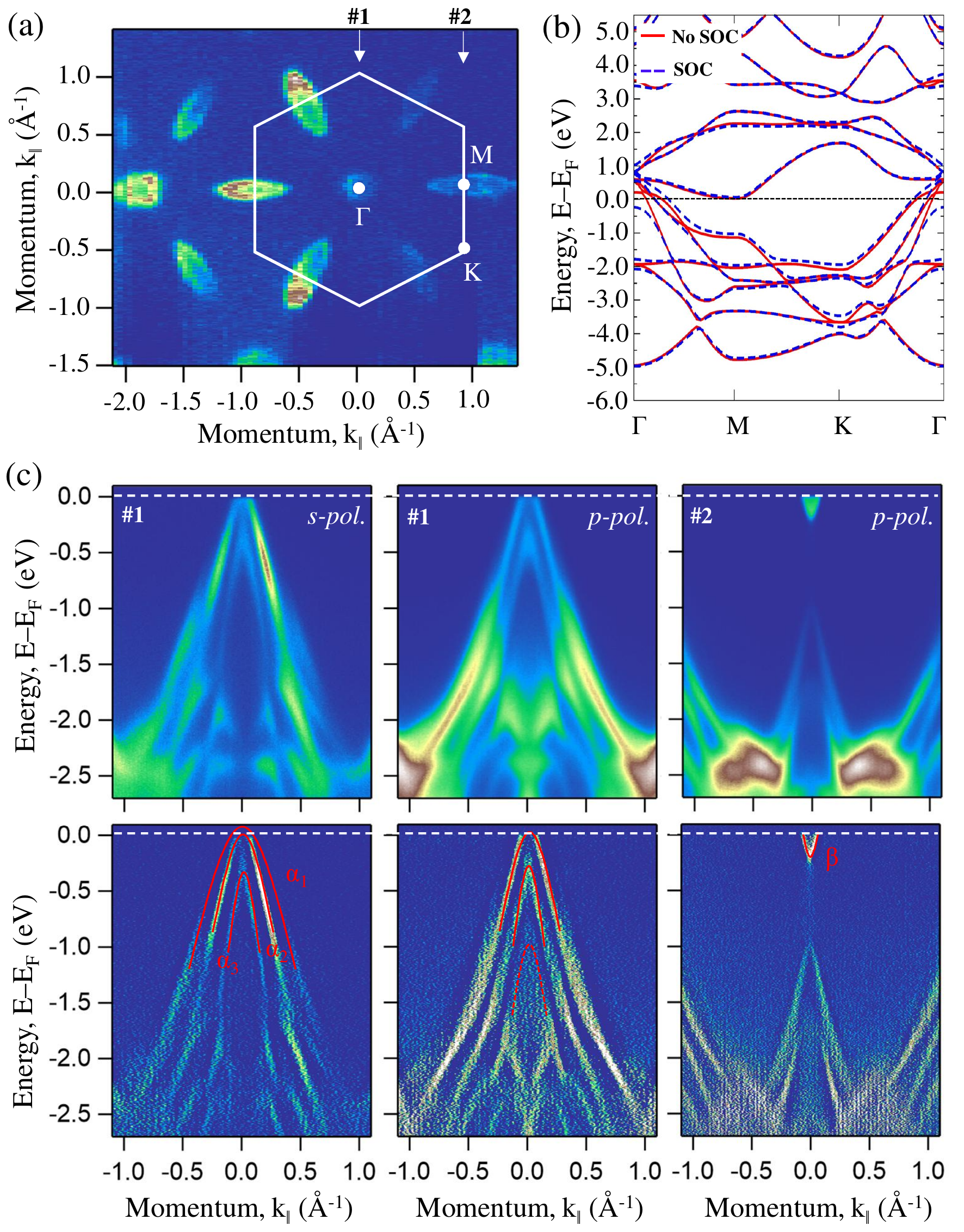}
  \caption{In-plane electronic structure of ZrTe$_{2}$. (a) Fermi surface map measured using $p$ polarized light with a photon energy h$\nu$ = 100 eV. (b) Energy distribution maps (EDMs) taken along the cuts \#1 and \#2 as shown in (a). (c) Second derivatives of (b). In (c) the red-dashed curves are eye-guides showing the band dispersions. (d) Energy-momentum plot of ZrTe$_2$ calculated using density functional theory with (blue) and without (red) including spin-orbit coupling.}
  \label{2}
\end{figure}

Figure~\ref{1}(a) shows the trigonal crystal structure of ZrX$_{2}$.  Figure~\ref{1}(b) shows the projected Zr-chalcogen layer on the $ab$-plane where one can notice each Zr ion is octahedrally coordinated with six Te (Se) ions.  From the scanning electron microscope (SEM) measurements on ZrTe$_2$ and ZrSe$_2$ (not shown) we derived the chemical formulae of Zr$_{1.02}$Se$_{1.98}$ and Zr$_{1.1}$Te$_{1.9}$, respectively,  suggesting that there is a 2\% of Se (excess Zr) and a 10\% of Te (excess Zr) deficiency from the stoichiometric compositions. Such a chalcogen deficiency or excess Zr in these systems generally leads to excess electron carrier density. Nevertheless, from our EDX we noticed that the samples are homogeneous.  As shown in Figures ~\ref{1}(g) and ~\ref{1}(h), we performed X-ray photoemission spectroscopy (XPS) on ZrTe$_2$ and ZrSe$_2$, respectively. Most of the atomic core levels are assigned to their respective binding energies for ZrTe$_2$. However, for ZrSe$_2$, we noticed significant intensity peaks at binding energies of 52 eV and 50.6 eV, pointed by {+} and {$\star$}, respectively, that are not visible in ZrTe$_2$. These peaks are identified as the Iodine 4p$_{3/2}$ ({+}) and 4p$_{5/2}$ ($\star$) states. Note here that Iodine is used as transport agent for growing these crystals, hence, some of the Iodine atoms may have stuck in between the ZrSe$_2$ layers during the crystal growth process. But nevertheless,  as we discuss below presence of Iodine in ZrSe$_2$ has no net effect on its electronic structure. Apart from this, we do not find any other impurity peaks from the XPS of ZrSe$_2$ and as well from ZrTe$_2$.

\begin{figure*}
  \centering
  \includegraphics[width=0.98\textwidth]{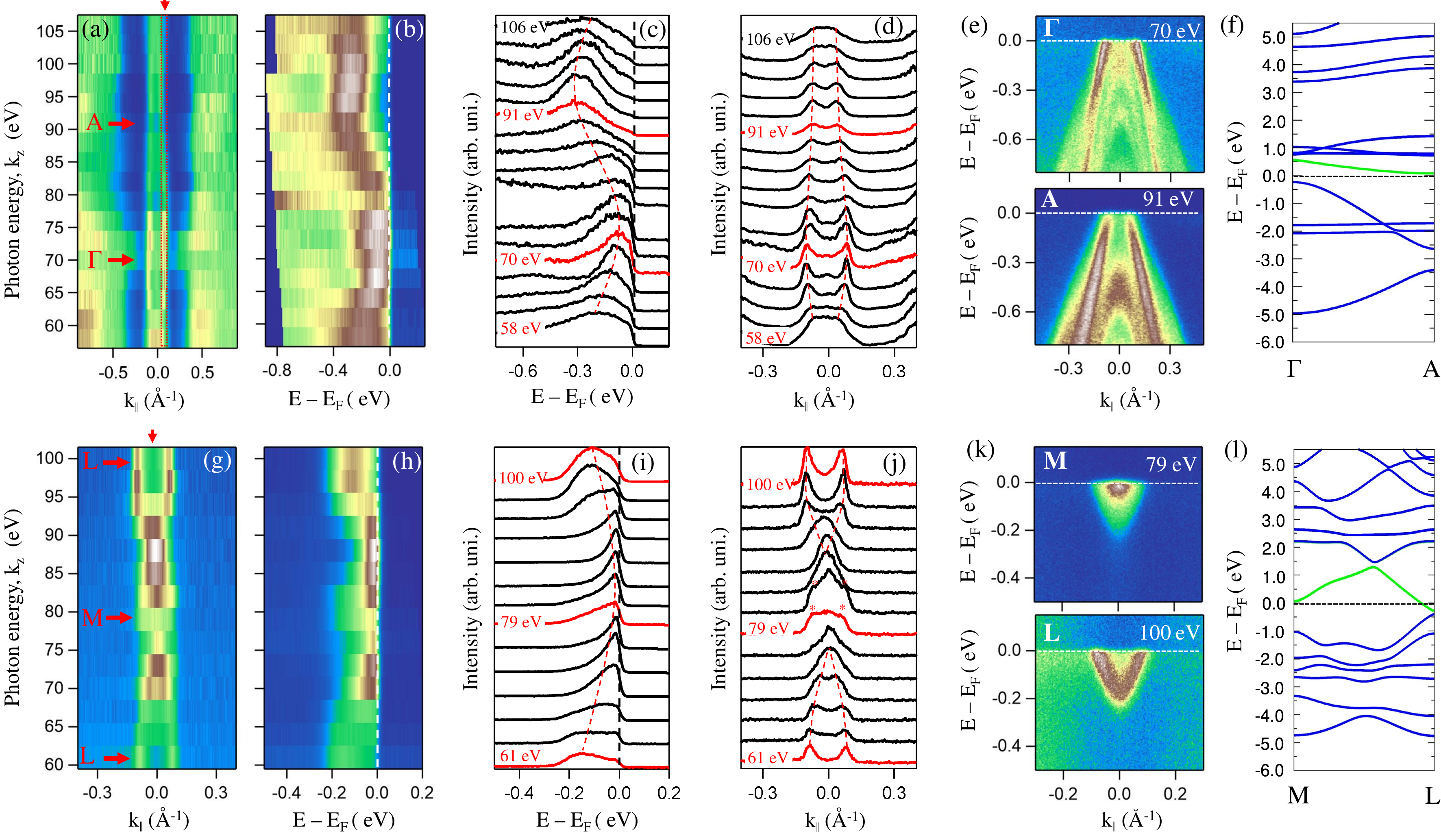}
  \caption{Out-of-plane ($k_z$) electronic structure of ZrTe$_{2}$. (a) k$_z$ Fermi surface map measured using $p$-polarized light in the $\Gamma M A L$ plane.  (b) EDM taken along the $\Gamma - A$ high symmetry line. (c) Photon energy dependent EDC curves extracted from (b). (d) Photon energy dependent MDC curves extracted from (a). (e) EDMs taken from $\Gamma$ and $A$ high symmetry points. (f) Energy-momentum plot from the calculations along the $\Gamma - A$ high symmetry line. (g) k$_z$ Fermi surface map measured using $p$-polarized light in the $M K H L$ plane. (h) EDM taken along the $M-L$ high symmetry line. (i) Photon energy dependent EDC curves extracted from (h). (i) Photon energy dependent MDC curves extracted from (g). (k) EDMs taken from $M$ and $L$ high symmetry points. (l) Energy-momentum plot from the calculations along the $M - L$ high symmetry line. }
  \label{3}
\end{figure*}

Figure~\ref{2} shows ARPES data of ZrTe$_{2}$. Fermi surface map of ZrTe$_{2}$ in the $k_x-k_y$ plane is shown in Figure~\ref{2}(a), measured with $p$-polarized light using a photon energy of h$\nu$=100 eV. The Fermi surface has hexagonal symmetry which is consistent with the hexagonal crystal structure.  As can be observed in Fig.~\ref{2}(a) the Fermi surface of ZrTe$_2$ consists of well disconnected Fermi pockets located at the $\Gamma$ and $M$ points. In order to understand nature of these Fermi pockets we have taken the energy distribution maps (EDMs) along the cuts \#1 and \#2 as shown in Fig.~\ref{2}(a). From the EDMs shown in Figure~\ref{2}(b) three holelike band dispersions, $\alpha_1$, $\alpha_2$, and $\alpha_3$,   are found at $\Gamma (A)$ and an electronlike band dispersion, $\beta$, is found at $M (L)$ high symmetry points. From the polarization dependent EDMs as shown in Fig.~\ref{2}(b), with $s$-polarized light we noticed all three hole pockets with reduced intensity for $\alpha_1$ and $\alpha_3$  and high intensity for $\alpha_2$,  whereas with $p$-polarized light we noticed predominantly two hole pockets ($\alpha_2$ and $\alpha_3$). Our orbital resolved band structure (see Figure~\ref{7} and Figure S1 in the supplementary) in the absence of SOC suggest that the hole pocket $\alpha_1$ is composed by Te $p_x$, the hole pocket $\alpha_2$ is composed by Te $p_y$, and the hole pocket $\alpha_3$ is composed by Te $p_z$  character. However, in the presence of SOC the orbital contribution to the hole pockets is more complex as can be observed in Fig.~\ref{7}. Thus, assigning orbital characters to the experimental band structure for ZrTe$_2$ is non-trivial as ARPES detects bands in the presence of SOC. Next, an high intense electron pocket ($\beta$) has been recorded using $p$-polarized light which is composed by the even parity orbital Zr $d_{z^2}$ character. Figure~\ref{2}(d) shows calculated band structure with and  without including spin-orbit coupling. As we can be see from Fig.~\ref{2}(d), three holelike and an electronlike band dispersions are predicted at $\Gamma (A)$  and $M (L)$ high symmetry points. Therefore, calculated band structure is quantitatively in agreement with the experimental data except that the Fermi level needs to be shifted $\approx$ 0.2 eV to high kinetic energy to match with the experimental Fermi level.




Despite being a layered system, we further performed  photon energy dependent ARPES measurements on ZrTe$_2$ to unravel the out-of-plane ($k_z$) band structure. Figure~\ref{3}(a) shows Fermi surface map of ZrTe$_2$ in the $\Gamma M L A$ plane measured with photon energies ranging from 58 to 106 eV in step of 3 eV using $p$-polarized light. Following the equation, $k_z~=~\sqrt{\frac{2m}{\hbar^2}}~\sqrt{V_0 + E_k\cos^2\theta}$,  and with an inner potential $V_0$ = 16 eV, we found that the photon energies 70 eV and 91 eV correspond to the high symmetry points of  $\Gamma$ and $A$, respectively. Figure~\ref{3}(b) shows EDM taken along the $\Gamma - A$ direction as shown by the down arrow in Fig.~\ref{3}(a) extracted by a momentum integration of 0.05 $\AA^{-1}$. From this EDM one can clearly notice that there is significant $k_z$ dispersion in going from $\Gamma$ to $A$ for the hole pocket which has a partial contribution from Zr $d_{z^2}$ along with Te $p_x (p_y)$ under SOC [see Fig.~\ref{7}].  This is further demonstrated using photon energy dependent EDCs as shown in Figure~\ref{3}(c). Figure~\ref{3}(d) shows photon energy momentum dispersive curves extracted from Fig.~\ref{3}(a). We estimated a Fermi vector of hole pocket $k_F \approx 0.09\AA^{-1}$ at the $\Gamma$-point, while a Fermi vector $k_F \approx 0.06\AA^{-1}$ is estimated at the $A$-point. This observation again confirms significant $k_z$ dispersion of the hole pocket.  Figure~\ref{3}(e) shows EDMs from $\Gamma$ (70 eV) and $A$ (91 eV) high symmetry points.  This observation is consistent with our DFT calculations along $\Gamma - A$ direction as shown in Fig.~\ref{3}(f), where one can see a significant $k_z$ dispersion for the hole pocket (green color band dispersion).

Similarly,  Figure~\ref{3}(g) shows $k_z$ Fermi surface map in the $M K L H$ plane measured with photon energies ranging from 61 eV to 100 eV in step of 3 eV using $p$-polarized light. From these data we found that the photon energy 79 eV detects bands from the $M$-point whereas the photon energies 61 and 100 eV detect the bands from the $L$-point. Figure~\ref{3}(h) shows EDM along the $M-L$ orientation and Figure~\ref{3}(i) shows photon energy dependent EDCs taken from Fig.~\ref{3}(h). Figure~\ref{3}(j) shows photon energy dependent MDC curves. We estimate a Fermi vector of $k_F \approx 0.08\AA^{-1}$ for the electron pocket at the $L$-point, while no Fermi vector for the electron pocket is found at the $M$-point. This observation is suggesting that the $k_z$ dispersion of the electron pocket composed by Zr d$_z^2$ orbital character is relatively stronger. Further, in Fig.~\ref{3}(j) we could identify a couple of peaks shown by $\star$ symbol on the MDC curves taken around the $M$ point. These peaks are caused by the reminiscent spectral intensity of the electron pocket from the $L$-point. This is because, in ARPES the photon energy dependent data generally suffers from the $k_z$ resolution due to vertical transition of the photoelectrons. Thus, the $k_z$ dependent ARPES data generally consists of overlapping states in the $k_z$ direction. Figure~\ref{3}(k) shows EDMs from $M$ (79 eV) and $L$ (100 eV) high symmetry points. This observation is consistent with our DFT calculations along $M - L$ direction as shown in Fig.~\ref{3}(l), where one can see a significant $k_z$ dispersion for the electron pocket (green color band dispersion).

\begin{figure}
  \centering
  \includegraphics [width=0.48\textwidth] {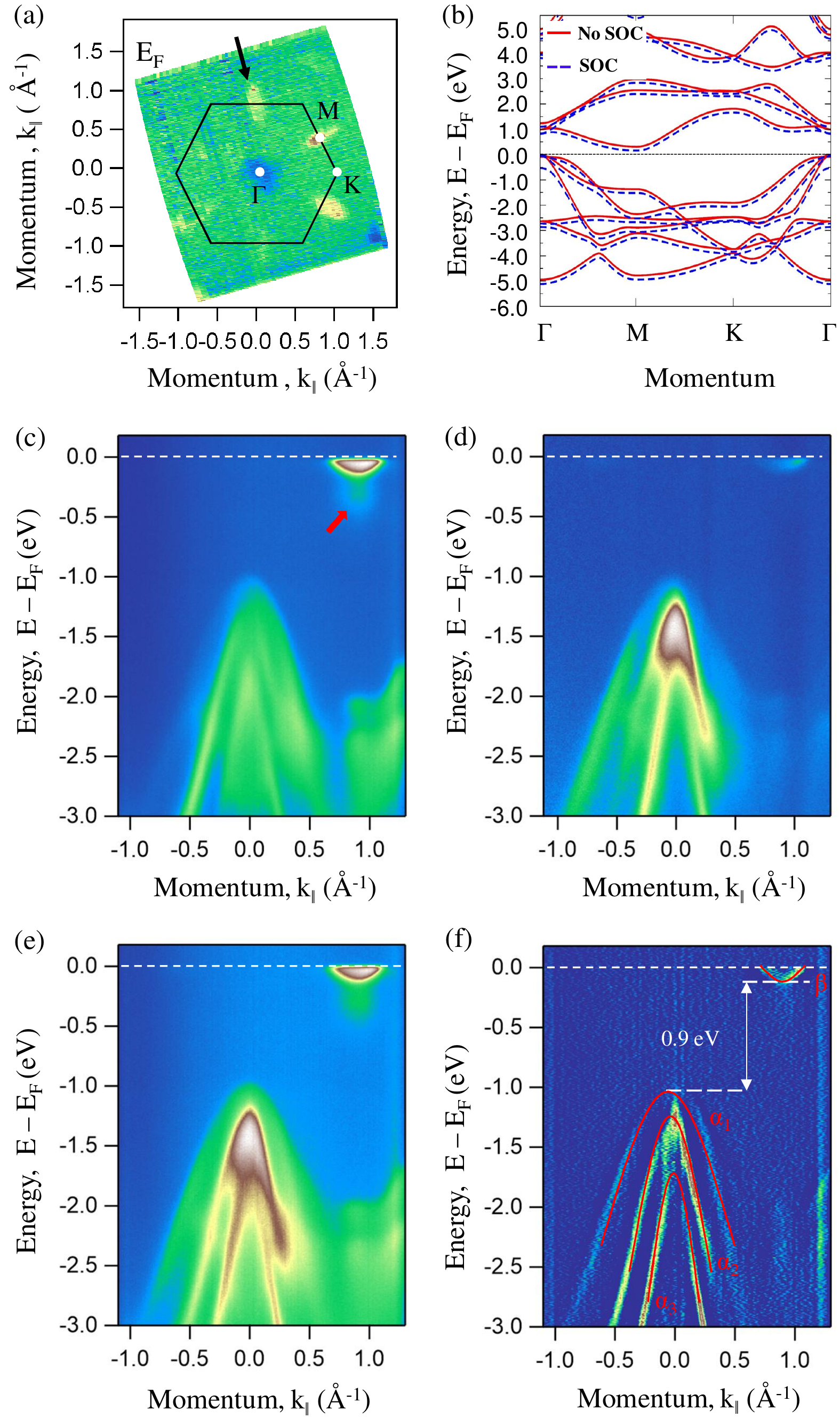}
  \caption{ARPES spectra of ZrSe$_2$. (a) Fermi surface map measured using $p$-polarized light with a photon energy of h$\nu$ = 100 eV. (b) Energy-momentum plot from the first-principles calculations. (c) EDM cut taken along the direction as shown in (a) measured using $p$-polarized light. (d) Same as (c) but measured using $s$-polarized light. (e) Sum of (c) and (d). (f) Second derivative of (e).}
  \label{4}
\end{figure}

Next, ARPES data of ZrSe$_2$ is shown in Figure~\ref{4}. Figure~\ref{4}(a) shows the Fermi surface map measured with a photon energy of h$\nu$ = 100 eV using $p$-polarized light in the $k_x-k_y$ plane. Figure~\ref{4} (b) shows energy-momentum plot of ZrSe$_2$ from DFT calculations. On the Fermi surface, we could identify six Fermi pockets located at six $M (L)$-points in a hexagonal symmetry. On the other hand, we did not find any spectral intensity at the $\Gamma (A)$ point. To further elucidate nature of the band dispersions at the $M (L)$ and $\Gamma (A)$ points, we measured energy distribution maps along the direction as shown in Fig.~\ref{4} (a). Figures~\ref{4} (c) and (d) are such EDMs measured with  photon energy of h$\nu$ = 100 eV using $p$- and $s$ polarized lights, respectively. From these EDMs it is clear that there are three well resolved holelike band dispersions, $\alpha_1$, $\alpha_2$, and $\alpha_3$, near the $\Gamma (A)$-point with the valance band top well below the Fermi level, at a binding energy of E$_B$ $\approx$ -1 eV. On the other hand, we found an electron like band dispersion, $\beta$,  crossing the Fermi level with a Fermi vector of $k_F$=0.06$\AA^{-1}$ at the $M (L)$-point. This electron pocket is almost disappeared when measured with $s$-polarized light as shown in Figure~\ref{4} (d).  The orbital resolved band structure of ZrSe$_2$ (see Figure~S2 in the supplementary) suggests that the holelike band dispersions are composed by Se $p_x$, $p_y$, and $p_z$ orbital characters, while the electronlike band near the $M (L)$-point is composed by the Zr $d_z^2$ orbital character. Thus, the electron pocket is more intense when measured using $p$-polarized light compared to $s$-polarized light as in the current measuring geometry near the Fermi level the  $p$-polarized light detects predominantly even parity orbital characters  like $d_z^2$, $p_x$ and $p_z$ whereas the $s$-polarized light detects $p_y$ character. Figure~\ref{4} (e) shows the sum of EDMs from Figs.~\ref{4} (c) and (d). Figure~\ref{4} (f) is the second derivative of Fig.~\ref{4} (e). From Figs.~\ref{4} (e) and (f) we estimated an indirect band gap of 0.9 eV between the top of holelike (valance) band, $\alpha_1$,  near the $\Gamma (A)$ point and the bottom of electronlike (conduction) band, $\beta$, near the $M (L)$-point. These observations are in good agreement with earlier ARPES studies on this system, i.e., ZrSe$_2$ is a semiconductor with an indirect band gap~\cite{Moustafa2009, Moustafa2013, Mleczko2017, Ghafari2018}. In addition, near the $M (L)$ point, from the ARPES data a bright spectral feature has been noticed at a binding energy of -0.3 eV [shown by red arrow in Fig.~\ref{4}(c)], which is not predicted from our DFT calculations. But similar spectral feature has been observed in an earlier ARPES report on ZrSe$_2$~\cite{Ghafari2018}.

\begin{figure}[b]
  \centering
  \includegraphics [width=0.49\textwidth] {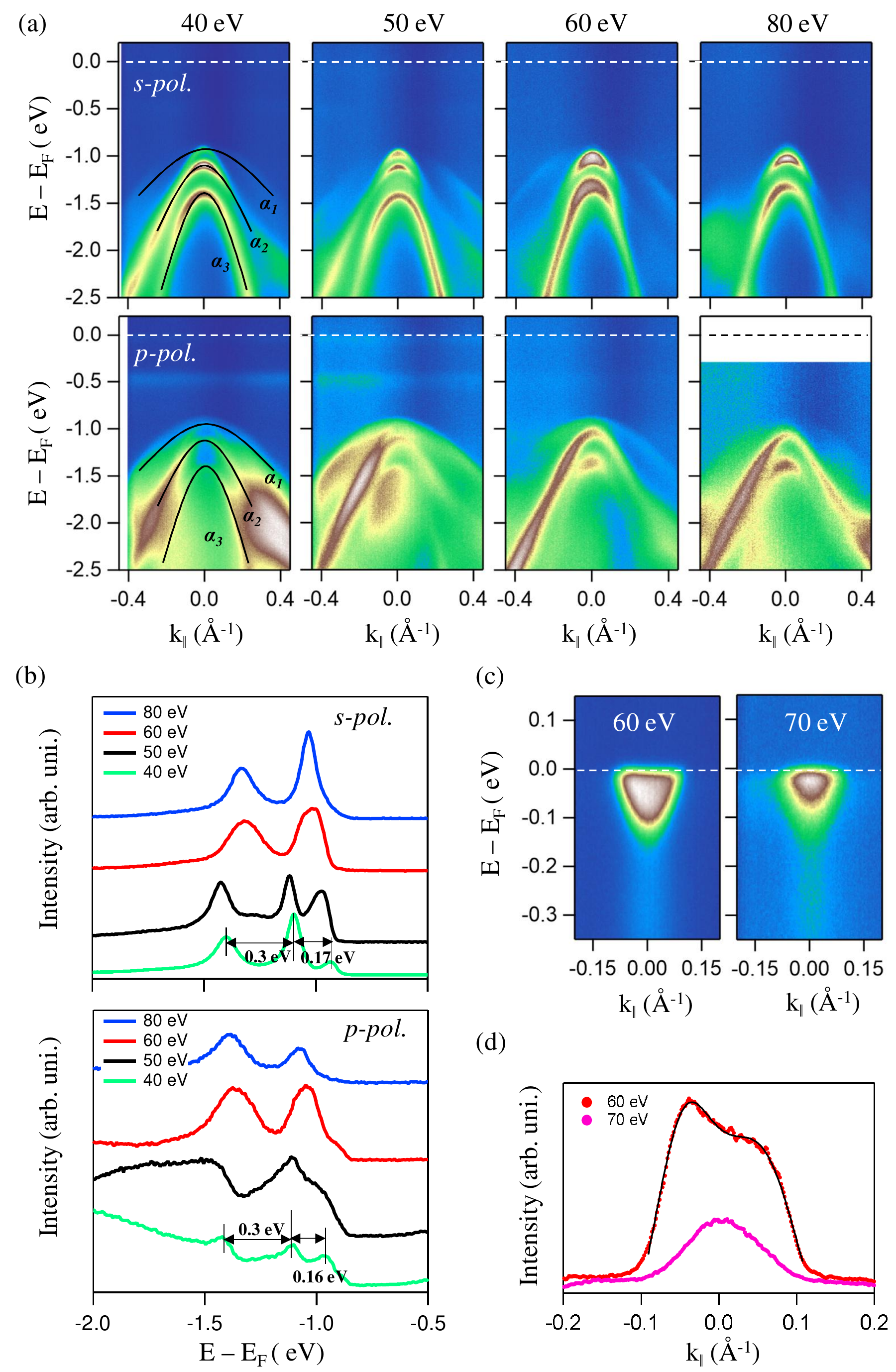}
  \caption{Photon energy dependent ARPES spectra of ZrSe$_{2}$. (a) Photon energy dependent EDMs measured using $s$-polarized (top panel) and $p$-polarized light (bottom panel) at the zone center. (b) Photon energy dependent EDCs from $s$- and $p$-polarized lights. (c) Photon energy dependent EDMs taken for the electron pocket. (d) Momentum dispersive curves extracted by integrating over 10 meV about the Fermi level,  taken from (c).}
  \label{5}
\end{figure}

 Top panels in Figure~\ref{5} (a) show photon energy dependent EDMs taken at the zone center using $s$-polarized light.  Similarly, bottom panels in figure~\ref{5} (a) show photon energy dependent EDMs taken at the zone center using $p$-polarized light. From the EMDs of both polarizations, we do notice three well resolved hole pockets at the zone centre. As discussed above, the $p$-polarized light should detect two holelike bands composed of the even parity orbitals $p_x$ and $p_z$ and the $s$-polarized light should detect one holelike band composed of odd parity orbital $p_y$. But since the crystal is not oriented in any high symmetry direction, the photoemission process does not follow the parity dependent selection rules and moreover the orbital contribution to the low-energy band structure of ZrSe$_2$ under SOC is more complex. Hence, we do see all three bands from both $p$- and $s$- polarized lights. Top panel in Figure~\ref{5} (b) shows photon energy dependent EDCs taken at $k_{\|}$=0 from the EDMs of $s$-polarized light. Similarly, bottom panel in  Figure~\ref{5} (b) shows photon energy dependent EDCs taken at $k_{\|}$=0 from the EDMs of $p$-polarized light. From the EDCs shown in Fig.~\ref{5} (b) one can clearly see that the three holelike bands are well separated. For instance from the EDCs of 40 eV in Fig.~\ref{5} (b), one peak has been observed at a binding energy of 1.4 eV, corresponding to the top of holelike band $\alpha_3$. Another two peaks have been observed at binding energies of  1.1 eV and 0.93 eV, corresponding to the top of holelike bands $\alpha_2$ and $\alpha_1$, respectively. Similar observations were made with 50 eV photon energy. However, when measured with 60 and 80 eV the band separation is not that clear due to either the matrix element effects or $k_z$ dispersion. Figure~\ref{5} (c) shows EDMs of the electron pocket measured using $p$-polarized light with 60 eV  and 70 eV. Figure~\ref{5} (d) shows momentum dispersive curves extracted near the Fermi level from the EDMs of Fig.~\ref{5} (c). From the MDC of 60 eV we estimated a Fermi vector of $k_F \approx 0.05 \AA^{-1}$, while the MDC of 70 eV shows no Fermi vector for the electron pocket. This suggest that the electron pocket composed of Zr $d_{z^2}$ character has significant $k_z$ dispersion in ZrSe$_2$.

\begin{figure}
  \centering
  \includegraphics [width=0.49\textwidth] {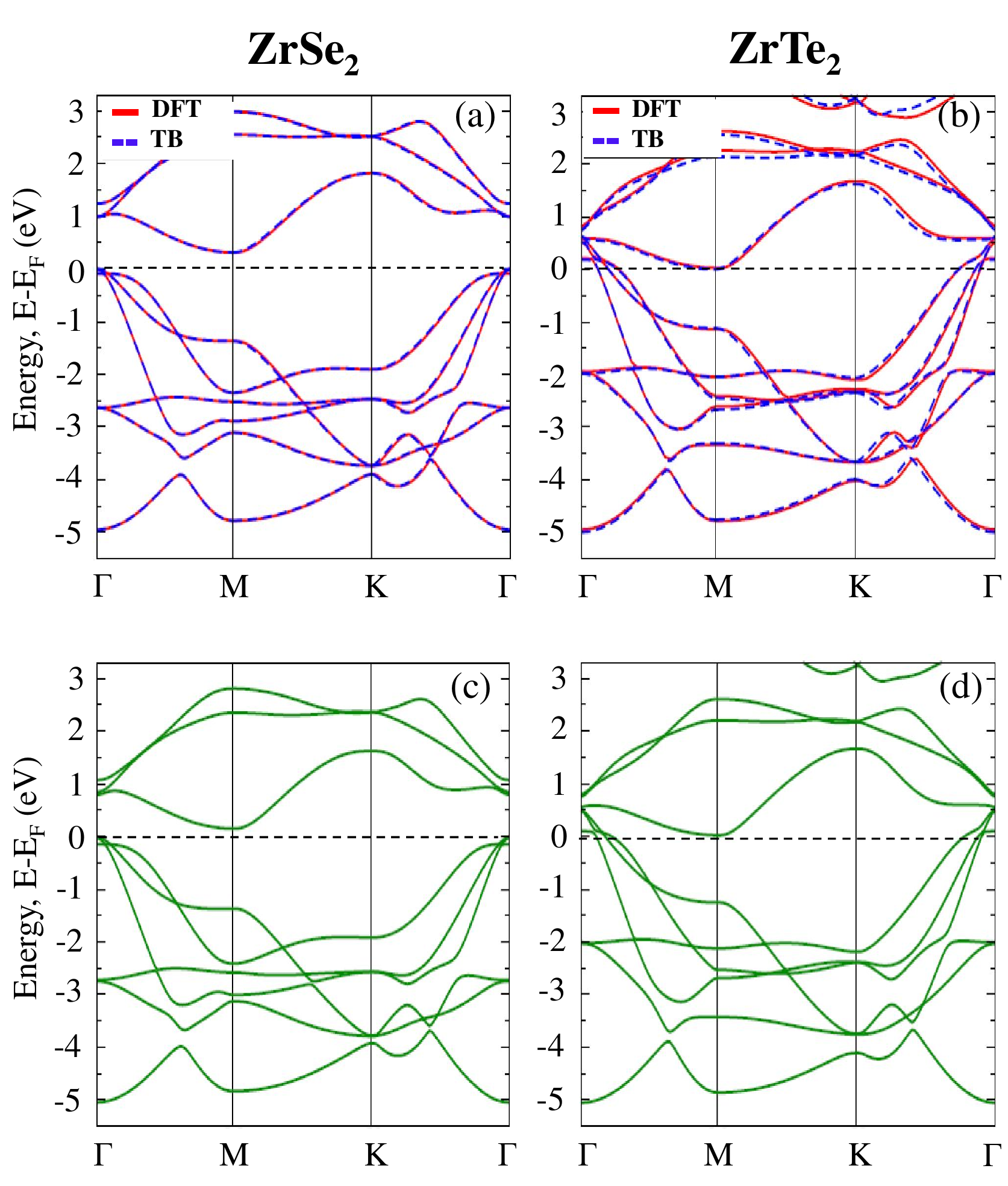}
  \caption{(a) and (b) Energy-momentum plots of ZrTe$_2$ and ZrSe$_2$ obtained form the DFT calculations and with overlapped bands derived from the tight-binding (TB) fittings, respectively. (c) Energy-momentum plot of ZrSe$_2$ derived from tight-binding fittings using the on-site energies of ZeTe$_2$. (d) Energy-momentum plot of ZrTe$_2$ derived from tight-binding fittings using the on-site energies of ZeSe$_2$. See supplementary information for further details on the on-site energies.}
  \label{6}
\end{figure}

\section{DISCUSSIONS}

Despite sharing the same trigonal crystal structure, the low-energy electronic structures of ZrTe$_2$ and ZrSe$_2$  are clearly differing from one another. ZrTe$_2$ has a metallic band structure with overlapping valance and conduction bands at the $\Gamma (A)$ point, composed by Te $p$ and Zr $d$ orbitals, respectively. On the other hand, ZrSe$_2$ has a semiconducting nature of the band structure with an indirect band gap of E$_g$=0.9 eV between $\Gamma (A)$ and $M (L)$ high symmetry points, with the valance and conduction bands composed of Se $p$ and Zr $d$ orbitals, respectively. In agreement with a previous ARPES report on ZrSe$_2$~\cite{Ghafari2018},  a tiny spectral feature near the $M (L)$-point at a binding of 0.3 eV has been observed. This feature is more intense when measured with $p$-polarized light than the $s$-polarized light, suggesting that it is composed by an even parity orbital. However, experimentally we are unable to pin point its band dispersion neither to electronlike nor to holelike. Moreover this feature is not predicted from the DFT calculations, thus cannot be of the intrinsic origin. As the XPS data of ZrSe$_2$ (see Fig. 1) shows Iodine peaks, perhaps, this Iodine impurity could be a plausible source of this strange spectral feature in ZrSe$_2$ than the back-folding of the bands from $M (L)$ to $\Gamma (A)$ as suggested by Ref.~\onlinecite{Ghafari2018}. Moreover, an earlier ARPES study on ZrSe$_2$ showed Fermi level lying in between valence and conduction band~\cite{Nikonov2018}, whereas from our study we found that the Fermi level is moved to the conduction band. This could be due to the Se deficiency in ZrSe$_2$ as observed from the EDAX measurements, which is giving rise to the net electron doping.

Spin-orbit coupling seems to be significantly effecting the low-energy electronic structure of these compounds. Especially, in ZrSe$_2$ the DFT calculations without SOC suggest two of the three hole pockets are degenerate, i.e.,  the two hole pockets composed by  $p_x$ and $p_y$ are degenerate while the third hole pocket composed by $p_z$ is well separated from them. On the other hand, the ARPES data of ZrSe$_2$ show three well separated hole pockets near the $\Gamma$ point. This can only reproduced in DFT calculations with the inclusion of SOC. In addition, the ARPES data of ZrSe$_2$ suggest a SOC splitting of hole pockets to as high as 0.17 eV. This observation is in very good agreement with the SOC split size of 0.16 eV for the same hole pockets reported by an earlier ARPES study on ZrSe2 ~\cite{Moustafa2013}. On the other hand, our DFT calculations with SOC only produced a SOC splitting size of 0.085 eV (see Fig.~2).

We further would like to show our DFT calculations [see  Figure~\ref{7}] on ZrTe$_2$ in which we identify a band inversion between Zr $d$ and Te $p$ states at the $\Gamma$ point. As discussed above,  in ZrSe$_2$ a clear separation is noticed between the valence and conduction bands composed by Se $p$ and Zr $d$ states, respectively. On the other hand,  in ZrTe$_2$,  the Zr $d_{z^2}$ state comes closer to the valance band and in fact touches the valance band at $\Gamma$. Further, in the presence of SOC we observe clearly a band inversion in ZrTe$_2$ in such a way that the Te $p_y$ state moves up to the conduction band, while the Zr $d_{z^2}$ moves further down to the valance band. As a result, a band gap of 10 meV is noticed [see inset in Fig.~\ref{7}(b)].

Next, coming to the discussion on the electronic structure of ZrTe$_2$, for the first time we report here the ARPES studies on bulk ZrTe$_2$. An earlier existing only ARPES study is on monolayer thickness ZrTe$_2$ film deposited on InAs (111),  suggesting a Dirac like linear bands at the $\Gamma (A)$ point~\cite{Tsipas2018}. Interestingly, our DFT calculations predict a band inversion between Zr $d$ and Te $p$ states at the $\Gamma (A)$ point, hinting at a possible non-trivial band structure in ZrTe$_2$ despite experimentally failing to detect any linear Dirac surface states. Furthermore, with the help of in-plane and out-of-plane  ARPES data and Luttinger's theorem~\cite{Luttinger1960, Oshikawa2000} we calculated the number of hole and electron carriers enclosed by the hole and electron Fermi pockets. Our calculation results to hole carrier density n$_h$ = 0.003 per unit cell (1.1 $\times$ 10$^{19}$/cm$^3$)and electron carriers n$_e$ = 0.0026 per unit cell (0.95 $\times$ 10$^{19}$/cm$^3$). Our estimated electron carrier density value is very close to the value of the order of 10$^{19}$/cm$^3$ reported for Cu$_{0.3}$ZrTe$_{1.2}$~\cite{Machado2017}, but much higher than the value of 5.07 $\times$ 10$^{15}$/cm$^3$ reported for ZrTe$_2$~\cite{Bhavsar2011}. In any case, carrier density reported in Ref.~\onlinecite{Bhavsar2011} is rather pointing ZrTe$_2$ to a semiconductor, while ZrTe$_2$ is widely known as a metal as also confirmed by the present study. Equal number of hole and electron carrier density as derived from our ARPES data in addition to the band inversion as seen from the DFT calculations is suggesting that ZrTe$_2$ is  a topological semimetal.

Finally, in order to understand the electronic structure changes from semiconducting to a topological semimetal in going from ZrSe$_2$ to ZrTe$_2$, we mapped the ab-initio band structure of these systems onto a tight-binding model. At first, the band structure of these systems obtained from tight binding model has been compared with the band structure derived from the DFT calculations. Thus, a good description of the ab-initio bands has been noticed as shown in the Figures~\ref{6} (a) and (c). As Se is more electronegative than Te, the Se $p$ states are expected to be deeper than the Te $p$ states. Indeed, our results for the extracted on-site energies show this trend (see supplementary information), with the charge transfer energy being smaller by 0.175 eV for ZrTe$_2$ than ZrSe$_2$. We then went on to question if it is this difference in the on-site of the energy of ZrSe$_2$ led it to a semiconductor, while ZrTe$_2$ is semimetallic. This is then addressed by setting up on-site energies in the tight-binding model for ZrSe$_2$ equal to those extracted for ZrTe$_2$. The derived results are shown in Fig.~\ref{6}(b). While a small change is found in the bandgap of ZrSe$_2$ compared to DFT calculations, the semiconducting nature retains. These observations suggest that the smaller Zr $d$ - Se $p$ bond length of 2.69 $\AA$ in ZrSe$_2$ leads to a larger $p-d$ hopping interaction strength in ZrSe$_2$ compared to ZrTe$_2$. Thus, the bandgap in ZrSe$_2$ can be understood in terms of a bonding-antibonding gap arising from the Zr $d$-Se $p$ hybridization. Here, the valence band states are the bonding states whereas the conduction bands are the antibonding states. So, an increase in the $p$-$d$ hopping strength is expected to increase the energy gap between valance and conduction bands. A similar analysis was carried out for ZrTe$_2$ using the on-site energies of ZrSe$_2$. As derived band dispersions are shown in Fig.~\ref{6}(d).  However, the band gap in ZrTe$_2$ was not open up. Hence, it is the smaller Zr $d$- Te $p$ hopping interaction strength associated with the longer Te $p$ - Zr $d$ bond of length of 2.89$\AA$ which does not allow ZrTe$_2$ to become semiconductor.

\section{CONCLUSIONS}
In conclusion, we systematically studied the low-energy electronic structure of ZrTe$_2$ and ZrSe$_2$ using angle resolved photoemission spectroscopy and density functional theory calculations. ARPES data of ZrTe$_2$ suggest several well disconnected hole and electron pockets at $\Gamma (A)$ and $M (L)$ points, respectively. From the ARPES data of ZrTe$_2$, we realize three holelike non-degenerate band dispersions near the $\Gamma (A)$ point and an electronlike band dispersion at the $M (L)$ point. The experimental observations are in good agreement with the DFT calculations. An equal number of hole and electron carrier density as estimated from our ARPES data suggest ZrTe$_2$ to a semimetal. In addition, the DFT calculations of ZrTe$_2$ in the presence of spin-orbit coupling suggest a band inversion between Te $p$ and Zr $d$ characters near $\Gamma$ point, hinting ZrTe$_2$ to a topological semimetal.  On the other hand,  from the Fermi surface topology of ZrSe$_2$ we observe only the electron pockets located at $M (L)$ point, while the hole pockets are noticed well below the Fermi level. Our studies on ZrSe$_2$ further suggested it to a semiconductor with an indirect band gap of 0.9 eV between $\Gamma (A)$ and $M (L)$ high symmetry points. Our calculations further demonstrate that the metal-chalcogen bond-length plays a vital role on the electronic structure changes of ZrX$_2$ ( X = Se and Te) in such a way that an electronic phase transition takes place from semiconducting to topological semimetal in going from ZrSe$_2$ to ZrTe$_2$.

\begin{figure}
  \centering
  \includegraphics [width=0.49\textwidth] {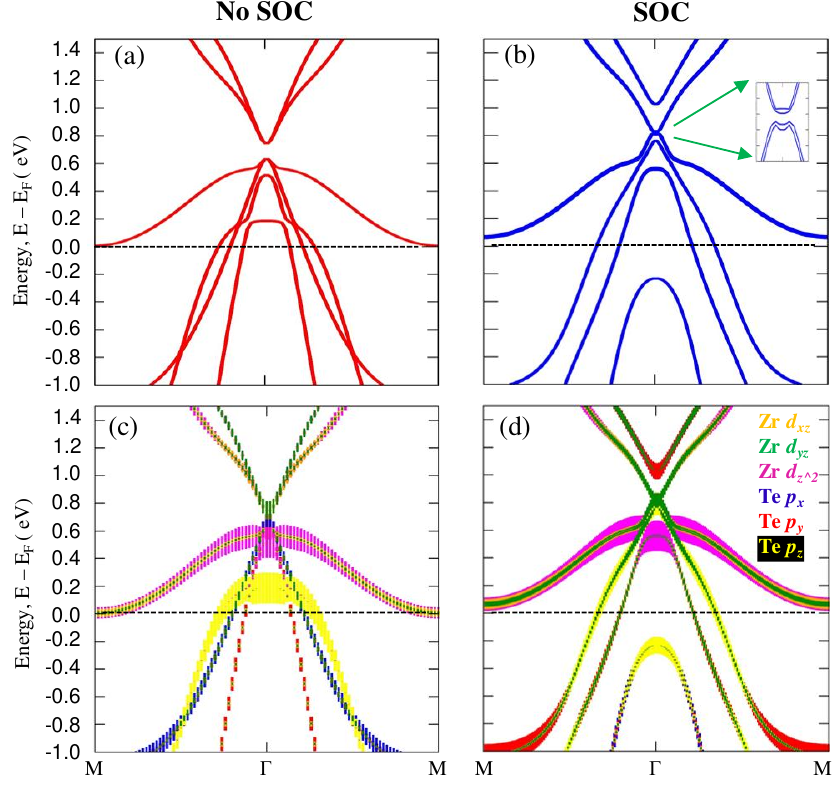}
  \caption{Calculated band structure of ZrTe$_2$ (a) without SOC and (b) with SOC. (c) and (d) are same as (a) and (b) but with orbital information. Inset in (b) shows a gap of 10 meV opening at the $\Gamma$ point due to the band inversion under the SOC.}
  \label{7}
\end{figure}

\section{ACKNOWLEDGEMENTS}
J.C. acknowledges the support from the Council of Scientific \& Industrial Research (CSIR), India. L.H acknowledges the financial by the Department of Science and Technology (DST), India (Grant No. SR/WOS-A/PM-33/2018 (G)) and Surjeet Singh for allowing the use of his crystal growth facilities. S.T. acknowledges support by the DST, India through the INSPIRE-Faculty program (Grant No. IFA14 PH-86). S. T. greatly acknowledges the financial support given by SNBNCBS through the Faculty Seed Grants program. This work was supported by the DFG under the Grant No. BO 1912/7-1.

\bibliography{ZrX2}

\end{document}